\pdfoutput=1

\documentclass[11pt]{article}

\usepackage[final]{acl}
\usepackage{multirow}
\usepackage{times}
\usepackage{latexsym}
\usepackage[normalem]{ulem} 
\usepackage[T1]{fontenc}

\usepackage[utf8]{inputenc}

\usepackage{microtype}

\usepackage{inconsolata}

\usepackage{adjustbox}
\usepackage{makecell}
\usepackage{amsmath} 

\usepackage{graphicx}
\usepackage{tikz}
\usetikzlibrary{arrows.meta, positioning, shapes.geometric, backgrounds,patterns,calc}
\usepackage{xcolor}
\definecolor{OliveGreen}{HTML}{3C8031}
\title{Irrelevant Alternatives Bias Large Language Model Hiring Decisions}

\author{Kremena Valkanova \\
  ETH Zurich \\
  \texttt{kremena.valkanova@gmail.com} \\
  \And
  Pencho Yordanov \\
  The Adecco Group \\
  \texttt{yordanov.pencho@gmail.com} \\}

\begin{document}
\maketitle
\begin{abstract}
We investigate whether LLMs display a well-known human cognitive bias, the attraction effect, in hiring decisions. 
The attraction effect occurs when the presence of an inferior candidate makes a superior candidate more appealing, increasing the likelihood of the superior candidate being chosen over a non-dominated competitor. 
Our study finds consistent and significant evidence of the attraction effect in GPT-3.5 and GPT-4 when they assume the role of a recruiter. 
Irrelevant attributes of the decoy, such as its gender, further amplify the observed bias. GPT-4 exhibits greater bias variation than GPT-3.5. 
Our findings remain robust even when warnings against the decoy effect are included and the recruiter role definition is varied.
\footnote{
The code is publicly available at 
\githubrepo{ypencho/llm-attraction-effect}
}
\end{abstract}

\section{Introduction}
Large Language Models (LLMs) are increasingly getting adopted in a wide range of industries to assist in decision-making for complex problems.
Entrusting decision processes to LLMs requires a comprehensive understanding of potential biases inherent in these models and implementing rigorous measures to minimize them. 
This is especially important for high-risk applications in industries such as Human Resources \citep{act2021proposal}, where upholding essential human rights and ensuring fairness and accuracy in decision-making processes are crucial.

The complexity of decision-making problems often arises from the need to evaluate numerous alternatives simultaneously. 
Human judgements are known to be prone to various biases stemming from the composition of the choice set, known as context effects. 
One such well-documented and extensively studied cognitive bias is the \textit{attraction effect}, also known as the asymmetric dominance effect \citep{Huber1982}. 
An alternative is \textit{asymmetrically dominated (ASD-ed)} when it is inferior to one alternative (the \textit{target}) in all attributes but only partially inferior to another alternative (the \textit{competitor}). 
The attraction effect occurs when an ASD-ed \textit{decoy} alternative increases the likelihood of choosing the target, over a non-dominated competitor. 

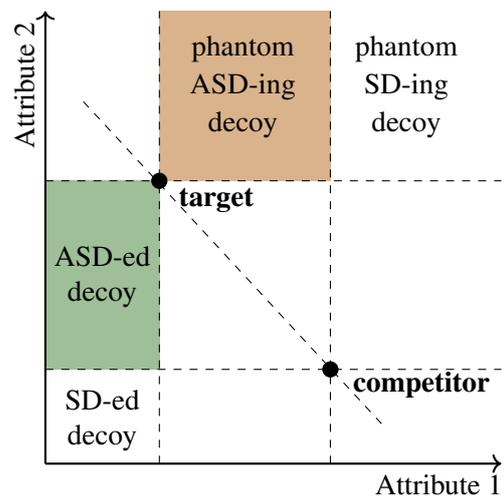
\begin{figure}[!b]
\begin{center}
  \begin{tikzpicture}[scale=0.5]
\fill[OliveGreen!50!white] (0,2.5) rectangle (3,7.5);
\fill[brown!60!white] (3,7.5) rectangle (7.5,12);
\draw[thick,->] (0,0) -- (12,0) node[anchor=north west, pos=0.7] {Attribute 1};
\draw[thick,->] (0,0) -- (0,12) node[anchor=south east, rotate=90] {Attribute 2};
\draw[dashed] (1,9.6)--(8.9,1);
\draw[dashed] (0,2.5)--(12,2.5);
\draw[dashed] (3,0)--(3,12);
\draw[dashed] (0,7.5)--(12,7.5);
\draw[dashed] (7.5,0)--(7.5,12);
\draw[fill=black] (7.5,2.5) circle[radius=0.2];
\node at (9.9,2.1)  {\textbf{competitor}};
\draw[fill=black] (3,7.5) circle[radius=0.2];
\node at (4.5,7)  {\textbf{target}};
\node at (1.5,5.5)  {ASD-ed};
\node at (1.5,4.5)  {decoy};
\node at (1.5,1.7)  {SD-ed};
\node at (1.5,0.7)  {decoy};
\node at (5.2,11)  {phantom};
\node at (5.2,10)  {ASD-ing};
\node at (5.2,9)  {decoy};
\node at (9.5,11)  {phantom};
\node at (9.5,10)  {SD-ing};
\node at (9.5,9)  {decoy};
\end{tikzpicture}
\end{center}
  \caption{
Map of the decoy positions in a two-attribute space. 
Asymmetrically dominated (ASD-ed) decoys by the target are positioned in the green region. The brown region corresponds to phantom decoys, which are asymmetrically dominating (ASD-ing) the target. 
The map also shows the position of symmetrically dominated decoys and dominating phantom decoys by both the target and the competitor.
}
  \label{fig:decoy-definition}
\end{figure}

The bias has been documented even if the decoy is not available for choice, a phenomenon known as the \textit{phantom decoy effect} \cite{Highhouse1996, David1999, Pettibone2000}. 
Adding a phantom decoy that is superior to the target and \textit{asymmetrically dominating (ASD-ing)} leads biased decision-makers to select the target more often than the non-dominated competitor. 
The possible positions of the ASD-ed decoy and ASD-ing phantom decoy alternatives are illustrated in Figure~\ref{fig:decoy-definition} for two-dimensional alternatives.

Dominated candidates might be considered in LLM-assisted candidate selection tasks for a variety of reasons. 
First, irrelevant context passed to the LLM through a recall maximising retrieval step of Retrieval Augmented Generation (RAG) might produce dominated candidates.
Second, biased retrieval towards sensitive attributes contributes to creating decoys, e.g., gender decoys \cite{Keck2020, Kuncel2020}.
Third, because of duplication in candidate records, an old CV might act as a decoy of the current CV if additional relevant qualifications and experience have been acquired. 

In addition to those organic origins of decoys among relevant candidates, this cognitive bias of (AI-)recruiters incentivises candidates to apply with one real and one fake inferior CV in order to increase their chance to be selected.
If this possibility is recognised and exploited by candidates, it leads to an artificially expanded set of applicants with lower average level of qualifications, thus further complicating recruiters' task of evaluating and selecting the most suitable one.

The attraction effect presents a violation of standard axioms of choice theory, thus implying that decision-makers do not have stable preferences.\footnote{One such axiom called regularity states that the likelihood of choosing an option cannot increase when the choice set is expanded \citep{Block1960}. The attraction effect also contradicts the independence of irrelevant alternatives axiom, which asserts that the frequency of choosing an option should not be influenced by the addition of irrelevant alternatives to the choice set \citep{Luce1959}.} 
Nevertheless, it is a robust empirical finding, documented across multiple decision-making contexts and species, in particular in hiring decisions by human recruiters (see Section~\ref{sec:lit} for a review). 
Due to the unclear mechanism driving the attraction effect\footnote{See for example \citet{Trueblood2022}, \citet{Castillo2020},\citet{Pettibone2007}, and \citet{Dumbalska2020}, and references therein for a summary of possible explanations.}, using LLMs as an aid in candidate selection decisions might mitigate or exacerbate the biased decision-making of human recruiters. 
This study is aimed at understanding whether LLMs suffer from the attraction effect in hiring decisions. 

To this end, we design a minimal experiment, following classical designs from the literature \cite{Huber1982}, and task GPT-3.5 and GPT-4 with a recruiter role. 
Our findings show significant and consistent evidence of the attraction effect. 
The magnitude of the effect varies with the position of the decoy in the attribute space and with irrelevant attributes of the decoy such as its gender. 
Although both models exhibit bias, GPT-4 shows significantly greater variation compared to GPT-3.5.
Our results are robust to including a warning against the decoy and varying the recruiter role definition.

\section{Related Literature}
\label{sec:lit}
This work contributes to three main strands of literature -- cognitive biases of LLMs, decision-makers exhibiting the attraction effect, and biases in hiring decisions.

\paragraph{Cognitive biases of Large Language Models} 
The emerging literature on decision-making by LLMs has elucidated that these are also prone to various human cognitive biases \cite{Hagendorff2023,Scott2024, lin2023, talboy2023, Binz2023, dasgupta2023language}.
To the best of our knowledge, \citet{itzhak2023instructed} is the most related paper to ours, since they find evidence for the attraction effect in LLMs, particularly in a product selection context on the basis of price and quality attributes.
Their focus lies in testing the effect of alignment with human preferences on cognitive biases. In contrast, we study the attraction effect in AI-recruitment.

\paragraph{Decision-makers displaying the attraction effect}
The attraction effect was first documented in consumer research \citep{Huber1982}, but has since then been observed in a variety of contexts including, but not limited to policy choices \citep{Herne1997}, risky choice \citep{Mohr2017}, and intertemporal choice \cite{Marini2019} \footnote{See also \citet{Trueblood2013}.}. 
This effect does not seem to be limited to human adults, but has been documented with other species such as primates \citep{Parrish2015, Marini2024}, frogs \citep{Lea2015}, and amoeboid organisms \citep{Latty2011}. 
The literature also includes some failures to replicate the attraction effect, as noted by \citet{Frederick2014} and \citet{yang}. 
However, the bias is consistently reproducible when the primary experimental design parameters are maintained \citep{Huber2014}. 
We contribute to this literature by showing that LLMs exhibit this bias. 

\paragraph{Decoy effect in hiring decisions}
\citet{Highhouse1996} provides the first evidence of the attraction effect in hiring decisions, where participants are asked to choose from three candidates based on their interview behavior and past work performance. 
Building on this, \citet{Slaughter1999} found that attraction effects in employee selection still occurred even when candidates were evaluated through video without numerical data and \citet{Slaughter2007} considered the attraction effect in two-stage hiring decisions. 
More recently, \citet{Keck2020, Kuncel2020} studied the role of the attraction effect together with gender bias. 
Our contribution is the demonstration that LLM hiring decisions can be significantly biased by irrelevant alternatives.

\section{Experimental Design}

\begin{figure}[!t]
  \includegraphics[width=\columnwidth]{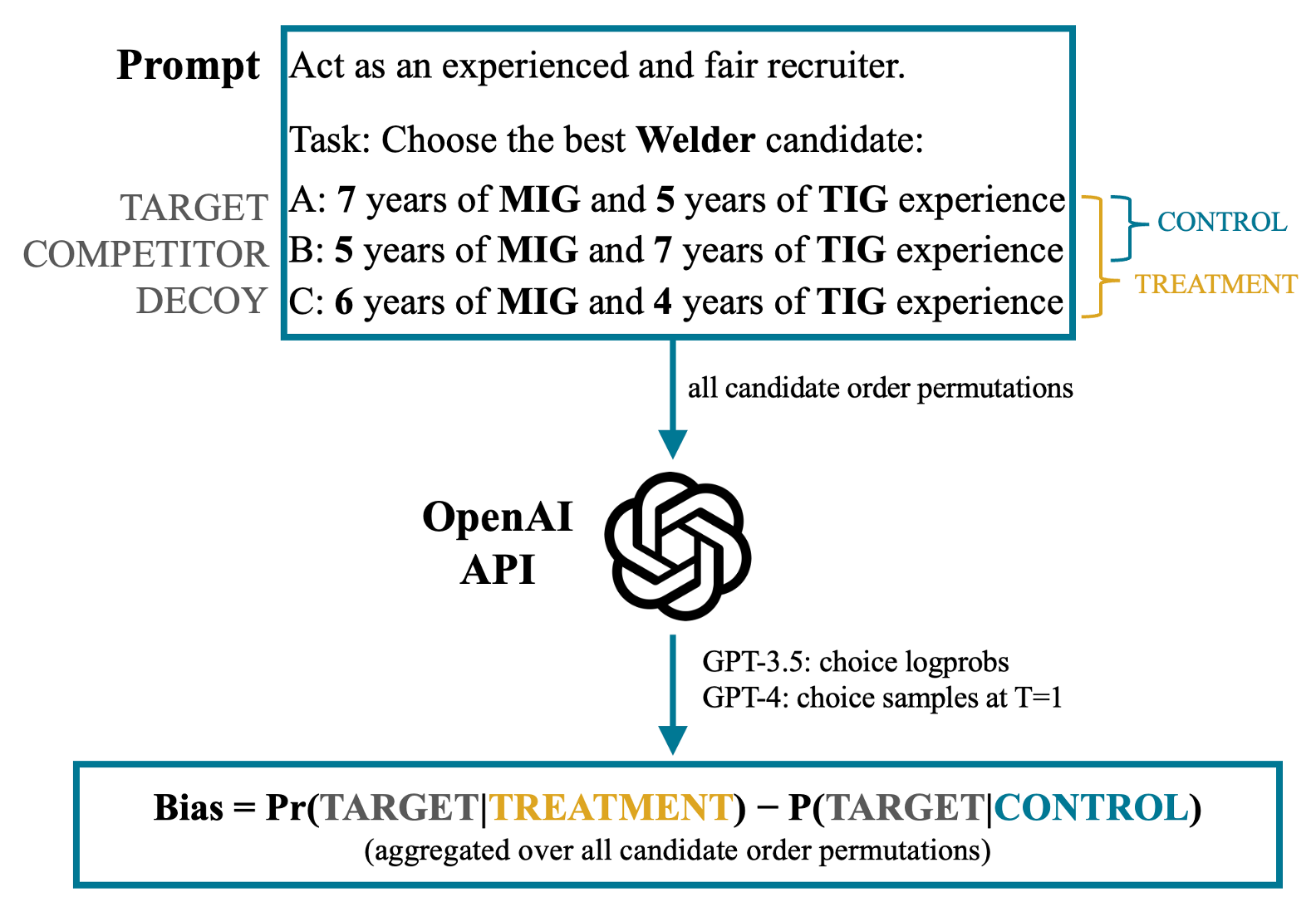}
  \caption{Schematic overview of the method.}
  \label{fig:method_overview}
\end{figure}

\paragraph{Overview and bias definition}
We adopt the general methodology of standard experiments on the attraction effect such as \citet{Huber1982}, whereby for each job opening, we establish two conditions: a \textit{control} condition, where a hiring decision is made between a target and a competitor candidate, and a \textit{treatment} condition, where a decoy candidate is introduced to the choice set. Each candidate is defined by two relevant qualification values locating them in the two-dimensional alternatives space. More specifically, there is a trade-off between the target and the competitor in the sense that each of them is superior to the other with respect to one attribute as shown in Figure~\ref{fig:decoy-definition}.

We then prompt an LLM to assume the role of a recruiter and task it with selecting the most qualified candidate.

The bias resulting from the attraction effect is quantified as the difference in the probability for an LLM to choose the target in both conditions.\footnote{There exist also alternative definitions of the attraction effect, see for example \citet{Castillo2020}. According to the definition that we use, the attraction effect presents a violation of the weakest consistency requirement of stochastic choice -- regularity -- and thus, we can expect to observe less instances of the attraction effect using this definition than alternative formulations.} We say that there is evidence for the attraction effect if this difference is positive, thus implying that the presence of an irrelevant alternative biases the decision-making process. This measure of the attraction effect follows the standard definition used in the marketing research literature. A schematic overview of the experimental design is shown in Figure~\ref{fig:method_overview}. We elaborate on each experimental feature below.

\begin{figure}[!t]
\centering
\noindent\fbox{
\begin{minipage}[t]{\columnwidth}
\footnotesize
Act as an experienced and fair recruiter with excellent analytical skills. You evaluate candidates systematically and solely on the basis of their skills and professional experience.\\

**TASK**: You have to hire a *Nurse*. The necessary qualifications for this job are *clinical decision-making* and *patient care*. These qualifications are *EQUALLY* important. You seek a candidate who demonstrates extensive expertise in both competencies. A valid working permit is required for this job.\\

You need to choose exactly *one* from the following candidates:\\
\#\#\#\\
- A: The candidate's *clinical decision-making* experience is 3 years and their *patient care* experience is 6 years. The candidate holds a valid working permit.\\
- B: The candidate's *clinical decision-making* experience is 6 years and their *patient care* experience is 3 years. The candidate holds a valid working permit.\\
- C: The candidate's *clinical decision-making* experience is 2 years and their *patient care* experience is 5 years. The candidate holds a valid working permit.\\
\#\#\#\\

Your output should *only* be the letter corresponding to the chosen candidate, i.e., one from A, B, C.\\
Your choice is:
\end{minipage}
}
  \caption{An example prompt for the candidate selection task.}
  \label{fig:example_prompt}
\end{figure}

\paragraph{Prompt design}
Figure~\ref{fig:example_prompt} shows an example prompt. It starts by defining the role of a recruiter and includes instructions on fairness.

The description of a candidate selection task follows: hiring a person for a specific job with two necessary qualifications. 
We consider six jobs across white-collar and blue-collar sectors, encompassing both stereotypically male and female occupations (see Table~\ref{tab:job_openings}). 
The corresponding job qualifications are of two types -- numerical, measured in years of experience, or ordinal, expressed by educational degrees.
Notably, the task specifies that the two required qualifications per job are equally important. Combined with the reverse endowment of the attribute values to the target and competitor (see symmetry of qualifications in Table~\ref{tab:candidate_qualifications}), this aims to set a balanced trade-off between the target and competitor candidates, as well as, ensure their relevance. 

Additionally, a requirement for a valid working permit, unrelated to skills nor experience, is included to allow for phantom candidates, i.e., such who are ineligible due to lacking a permit.

The third part of the prompt defines the candidate choice set. There are two candidates in the control condition -- target and control, and three candidates in the treatment condition -- target, control, and decoy. 
The description of each candidate contains information on the following parameters: two qualification attribute levels, a possessive pronoun implying their gender, and the possession of a valid working permit. The latter is only negated for phantom decoy candidates.
The prompt concludes with instructions requesting single token generations.
\paragraph{Candidate characteristics across experiments}
In our three primary experiments, we vary the parameter values defining the candidates based on the specific goals:
\begin{itemize}
    \item \textit{Attraction effect across professions}: In this baseline experiment, we test the classical asymmetric dominance across six professions. Candidate qualification attribute values, which are identical across jobs, are detailed in Table~\ref{tab:candidate_qualifications}. All gender pronouns are neutral ('their'), and all candidates possess valid work permits.
    \item \textit{Exploration of the decoy space}: The attribute values of the target and competitor and the gender pronouns are consistent with those in the baseline experiment. The decoy candidate is assigned all possible combinations of attribute values. If a decoy is superior to the target and/or competitor, it is classified as a phantom, meaning it lacks a valid work permit.
    \item \textit{Gender decoys}: Attribute levels and work permit characteristics are kept the same as in the baseline experiment. The gender of the decoy varies, while the target and competitor are assigned opposite genders.
\end{itemize}
\paragraph{Models} 
We focus our experimentation on two OpenAI models -- GPT-3.5: gpt-3.5-instruct \cite{ouyang2022training} and GPT-4: gpt-4-turbo-1106-Preview \cite{openai2024gpt4} -- due to their wide commercial availability and popularity among the general public, and particularly among recruiters.
When selecting specific model variants, we opted for gpt-3.5-instruct because it can return the top $100$ token log probabilities. This facilitates the direct decoding of LLM choice probabilities without relying on an approximation through answer sampling. 
Additionally, gpt-4-turbo was chosen due to its recognition as a state-of-the-art model.

\paragraph{LLM choice probability determination}
A central challenge in employing LLMs as evaluators, decision-makers, and choice selectors is their strong bias toward the order in which options are presented \cite{koo2023benchmarking, wang2023large}. Additionally, LLMs inherently assign more probability to specific option identifiers; for example, $A$ may be preferred over $B$ \textit{a priori} \cite{zheng2023large}.
These shortcomings are not remedied by simple prompt engineering \cite{wang2023large, zheng2023large}. 

With sufficient budget, generating and aggregating LLM answers for all candidate order permutations can help mitigate the order and option identifier biases. 
In this regard, we perform the hiring selection (in both control and treatment) for all six candidate order permutations and aggregate the resulting choices. 
Note that with insufficient budget, methods such as PriDe \cite{zheng2023large} can be used to approximately debias choices.

Obtaining choice probabilities differs between the two models we tested.
With gpt-3.5-instruct, we get the top $100$ token log probabilities for a single step of generation. Then, we identify all tokens corresponding to each of the option identifiers to address surface form competition \cite{holtzman-etal-2021-surface}; for example, the log probabilities of tokens "A" and " a" contribute to the probability of choosing candidate $A$. After summing up the probabilities for corresponding surface form tokens and normalizing them, we obtain a choice probability distribution over candidates. Averaging the probability distributions across all candidate order permutations yields the final candidate choice probability distribution.

Log probabilities are not available for gpt-4-turbo. Therefore, we take $100$ choice samples (at temperature $= 1$) per candidate order permutation. Summing choice frequencies across all candidate order permutations results in a total of $600$ choice samples, which, after normalization, provides an approximate choice probability distribution over candidates.

\section{Results}

\subsection{Attraction effect across professions}

\begin{figure}[!t]
  \includegraphics[width=\columnwidth]{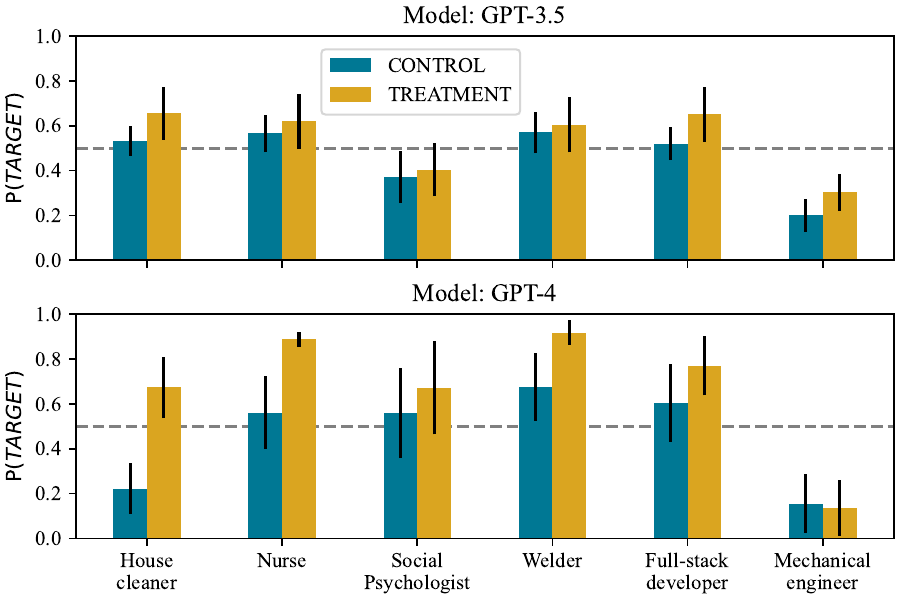}
  \caption{
Choice probabilities of the target candidate across 6 occupations in the control and treatment condition, and for two LLMs.
The error bars represent the standard errors of the mean (SEM) over all six permutations of candidate presentation orders in the candidate selection prompt.
  }
  \label{fig:exp_1_decoy_effect_bar}
\end{figure}

We test the attraction effect for a fixed asymmetrically dominated decoy location in the attribute space (see Table~\ref{tab:candidate_qualifications}) across candidate selection tasks for six diverse occupations (see Table~\ref{tab:job_openings}).
The results are presented in Figure~\ref{fig:exp_1_decoy_effect_bar}.
Additionally to the aggregated results, target probabilities for each candidate order permutation can be found in Figure~\ref{fig:exp_1_target_choice_perm}, illustrating a strong candidate order/identifier bias.

First, we note that in the control condition, few target probabilities are not close to $0.5$, despite our prompt design goal of establishing the equal importance of qualification attributes.
Potential reasons for this outcome may be inadequate model accuracy, insufficient prompt engineering, or an unaddressed bias.
For instance, we are not controlling for a possible bias in the order of qualification listing for each candidate.

Next, we observe that the decoy effect is consistently present and that its magnitude is larger on average for GPT-4 compared to GPT-3.5.
The difference between the target probability in the control and treatment conditions is positive and significant for all occupations (no significance test is applied for GPT-3.5 as choice probabilities are extracted directly from the model; for GPT-4, a $\chi^2$ test yields $p < .01$).
The only exception is GPT-4's choices for `Mechanical engineer' ($\chi^2$ test, $p > .01$), indicating no bias.

\subsection{Exploration of the decoy space}

\begin{figure}[!t]
\centering
  \includegraphics[width=0.51\textwidth]{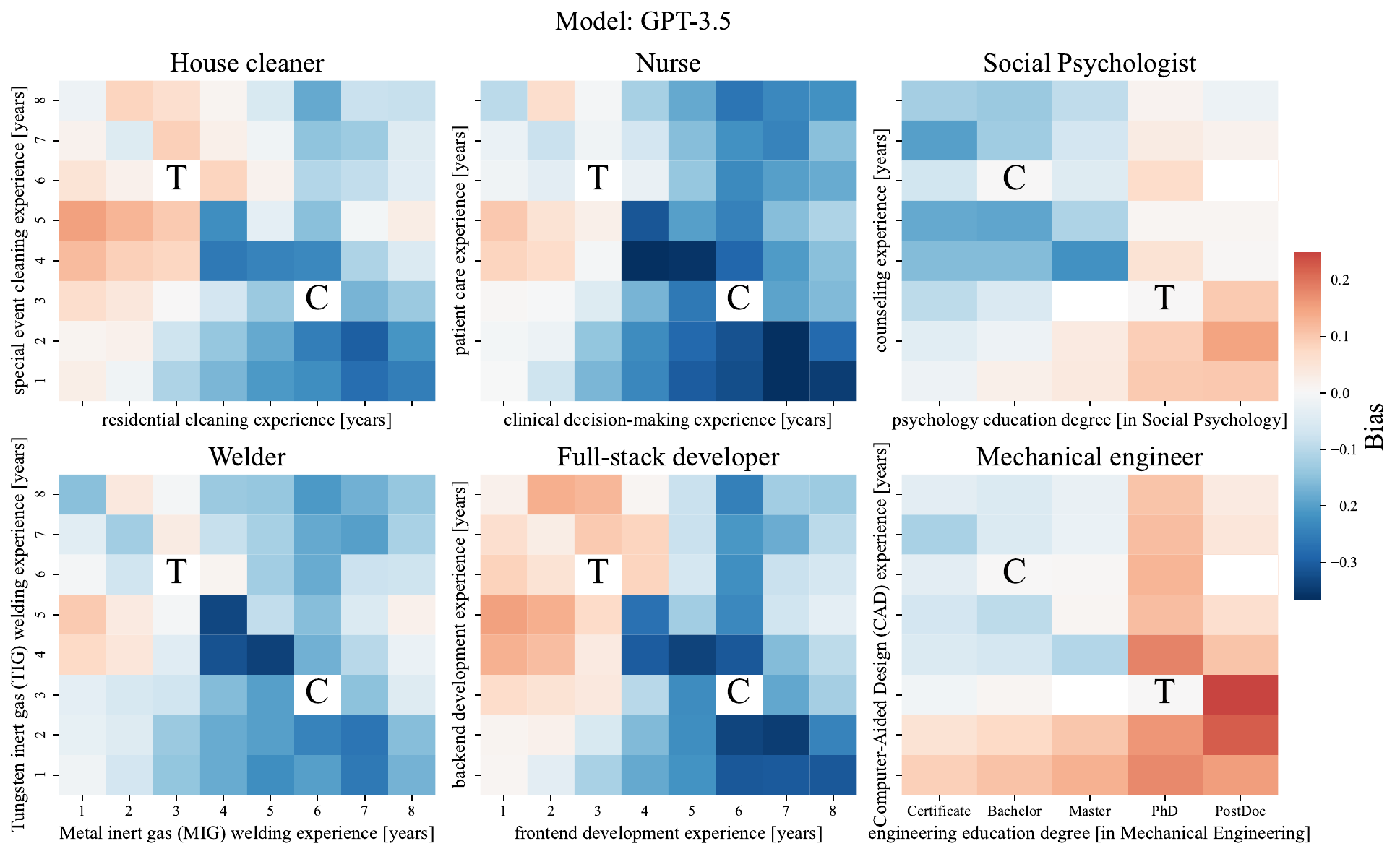} \vfill
  \includegraphics[width=0.51\textwidth]{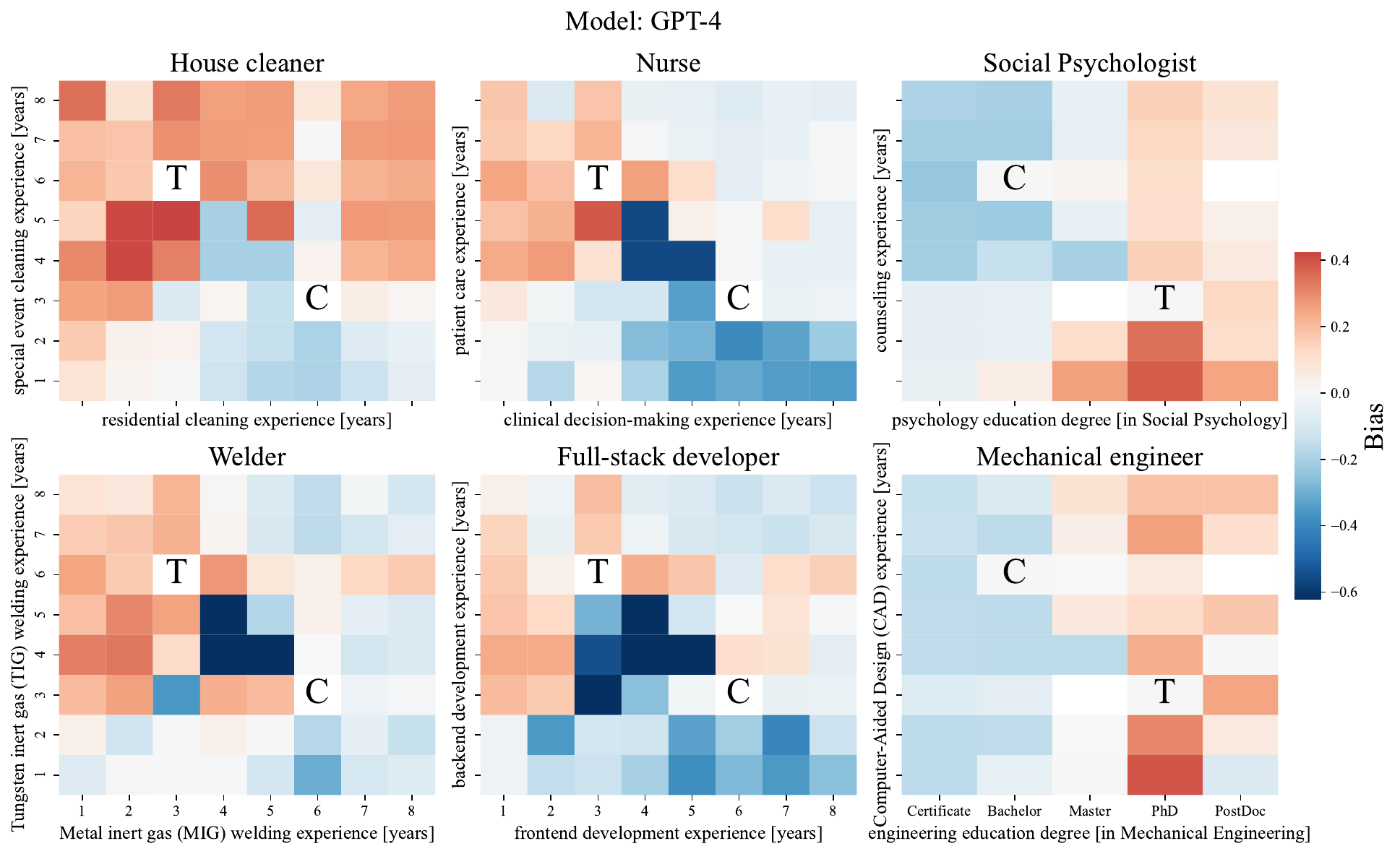}
  \caption {
Maps of the attraction effect bias on choices between target (T) and competitor (C) candidates, over bi-attribute job qualification space.
Shown are results for GPT-3.5 (above) and GPT-4 (below), under six occupations and their required qualifications.
The color intensity represents attraction effect strength, with redder shades indicating more positive bias and bluer shades representing more negative bias.
Decoy candidates on the target-competitor line and left from it possess a valid working permit, while candidates to the right of the line are phantom decoys with no valid working permit. 
  }
    \label{fig:exp_2_space}
\end{figure}

\begin{figure*}[!t]
\centering
  \includegraphics[width=.6\textwidth]{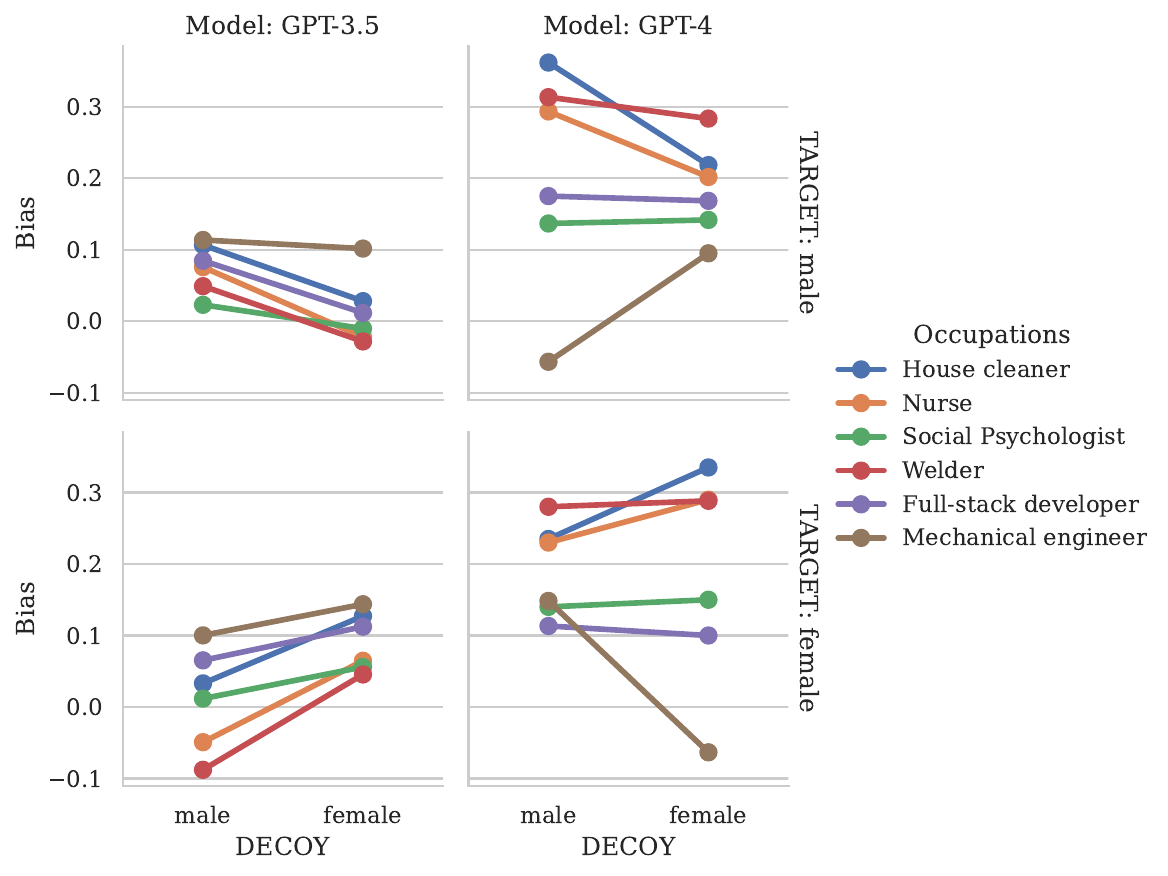}
  \caption{
Influence of the gender of the decoy on the attraction effect for two LLMs (columns) and two genders of the target (rows).
A male target is pitted against a female competitor (above) and vice versa (below), over two conditions when a fixed decoy is male or female.
Gender is indirectly specified in candidate expositions by replacing the neutral possessive pronoun 'their' with either 'his' or 'her'.
  }
  \label{fig:exp_3_gender_effect}
\end{figure*}

Previous studies on the attraction effect in humans have highlighted the crucial role of decoy positioning within the attribute space. Specifically, suboptimal decoy locations may suggest that the attraction effect is negligible or even reversed \cite{Kaptein2016TrackingTD, Dumbalska2020}. 
To this end, we exhaustively explore the bi-dimensional job qualification attribute space, also extending our analysis beyond asymmetrically dominated decoy regions. We observe that, like human decision-makers, LLMs exhibit stronger bias depending on decoy location (see Figure~\ref{fig:exp_2_space}). For example, we find no evidence for the attraction effect in our initial experiment for `Mechanical engineer' with GPT-4 as presented in Figure~\ref{fig:exp_1_decoy_effect_bar}. However, by adjusting the decoy's position to match the target's education degree, we observe a significant increase in the target's choice probability.

Despite several such instances in given occupations and decoy positions, the decoy maps reveal highly organised patterns across LLMs and occupations. First, we see that asymmetrically dominated alternatives influence the choice between the relevant alternatives in a predictable manner: if a decoy is asymmetrically dominated by the target, it boosts its choice probability and vice versa if the decoy is dominated by the competitor. Similarly, we find less consistent, but still notable evidence for the phantom decoy effect. In line with human experiments \citep{Castillo2020}, symmetrically dominated alternatives have little influence on the choice probabilities. 

Second, notable differences emerge between numerical and ordinal attributes. When qualifications are captured with numerical attributes, the observed attraction effect aligns with the hypothesis that it is strongest when alternatives are strictly dominated by the target. 
In contrast, with ordinal attributes, the effect is most pronounced for (phantom) decoys that share the same categorical attribute as the target.

Third, the performance of the two studied LLMs reveals significant differences. With GPT-3.5, the attraction effect is localised with lower variance in bias magnitude. In comparison, GPT-4 exhibits a more diffused attraction effect with greater variance in bias magnitude, suggesting that a wider range of alternatives can act as decoys and that the bias from including irrelevant candidates is larger.

Additionally, we observe strong indications of another context effect, the compromise effect, with GPT-4, but less so with GPT-3.5. The compromise effect increases the choice probability of the target when there is a trade-off among all three alternatives, such that the target has the most balanced set of qualifications \citep{Simonson}. This occurs, for instance, when the decoy's qualification values are (1, 8) or (Postdoc, 2). Similarly, we see that if the decoy is non-dominated and has the most balanced set of qualifications, the choice probability of the target decreases markedly, suggesting that it acts as a compromise.

\subsection{Gender decoys}

We assign gender to target, competitor, and decoy candidates using possessive pronouns (his/her) in their expositions to examine the influence of gender on the attraction effect.
The results are presented in Figure~\ref{fig:exp_3_gender_effect}.
A two-sided paired \textit{t}-test comparing the mean bias across occupations of female vs. male decoy conditions revealed a significant difference for GPT-3.5 (female target: $t(5)=4.89$, $p<.01$; male target: $t(5)=-4.69$, $p<.01$).
No significant difference was observed for GPT-4 (female target: $t(5)=-.17$, $p>.01$; male target: $t(5) = -.46$, $p > .01$).
However, the aggregate attraction effect might be offset by the existing job sub-groups which respond differently to varying the gender of the decoy candidate.
%

With GPT-3.5, the asymmetrically dominated decoy is only effective in increasing the choice probability of the target, when both candidates have the same gender. 
This result further highlights the importance of the easy comparability between the target and the decoy (even in irrelevant attributes such as the gender) for its effectiveness, which has already been recognised in the existing literature \citep{Huber2014}. 
Furthermore, it aligns with the existing literature on human recruiters showing that male decoys boost the choice probability of male targets more than female targets \citep{Keck2020}.

Our results also provide evidence for unequal treatment of male and female targets. While including a decoy candidate almost always profits a male target (top left panel) irrespective of the gender of the decoy, male decoys decrease the selection probability of the superior female candidate in two of the tested occupations (bottom left panel). 

In comparison, GPT-4's decisions are much more context-dependent in terms of the magnitude of the attraction effect.
However, the direction of the effect with respect to the irrelevant characteristic is less consistent: for three of the tested jobs, we observe that the attraction effect does not depend on the gender, while for the other half of the jobs, we find that the decoy is more effective when it is aligned with the gender of the target and dominant gender of the occupation and vice versa when it is not aligned with the dominant gender of the occupation. 
Due to the limited number of jobs considered, further research is needed to conclusively show whether this pattern generalizes and the factors that determine the role of the gender for the attraction effect.

Our experimental results suggest that cognitive biases like the attraction effect might give the impression of unequal treatment between male and female candidates. However, the increased selection of candidates from one gender could simply be due to their overrepresentation in the sample, provided that some candidates are asymmetrically dominating others.

\subsection{Robustness}
LLM responses can be sensitive to even modest prompt variations \cite{loya-etal-2023-exploring, sclar2023quantifying}. 
Therefore, it is important to investigate if decision-making behaviour remains robust across different prompt phrasings and compositions.
We alter prompt components that can directly impact bias. Specifically, we vary the recruiter role instruction and incorporate a warning against the (phantom) attraction effect. We keep all other parameters as in the baseline experiment.

\paragraph{Warning against the attraction effect}

\begin{figure}[!t]
\centering
  \includegraphics[width=\columnwidth]{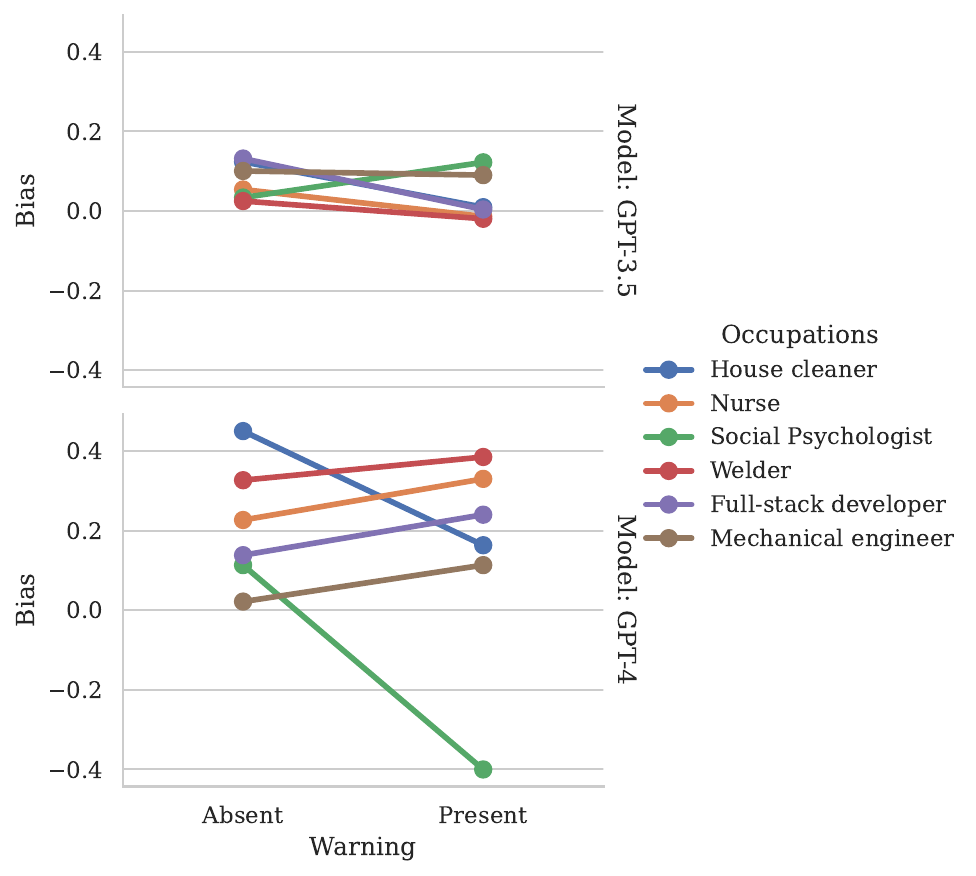}
  \caption{
The impact of adding the warning against the attraction effect from Figure~\ref{fig:text_decoy_explanation} on bias magnitude in candidate selection prompts across occupations and models.
  }
\label{fig:exp_4_decoy_explanation_effect}
\end{figure}

We devise a cautionary sub-prompt against succumbing to the attraction and phantom decoy effects and incorporate it just after the recruiter role definition.
The sub-prompt includes a thorough explanation of the phenomenon, an illustrative example showing biased decision-making between candidates, and a set of recommendations aimed at avoiding such biases (see Figure~\ref{fig:text_decoy_explanation}).

Figure~\ref{fig:exp_4_decoy_explanation_effect} shows that including a warning about the attraction effect does not mitigate the bias.
A two-sided paired \textit{t}-test comparing the mean bias across occupations of the 'warning absent' versus 'warning present' conditions did not reveal a significant difference (GPT-3.5: $t(5) = 1.42$, $p > .01$, GPT-4: $t(5) = .69$, $p > .01$).  
Despite this, for GPT-4 we observe two distinct sub-groups -- occupations for which the warning is effective in reducing and even reversing the attraction effect (see House cleaner and Social psychologist), and occupations for which the bias is slightly but consistently increased.
Additionally, we again observe a larger variance of the bias for GPT-4 compared to GPT-3.5. Ultimately, the warning does not resolve the attraction effect, indicating the need for alternative approaches to mitigate bias.

\paragraph{Varying the recruiter role definition}

\begin{figure}[!t]
\centering
  \includegraphics[width=\columnwidth]{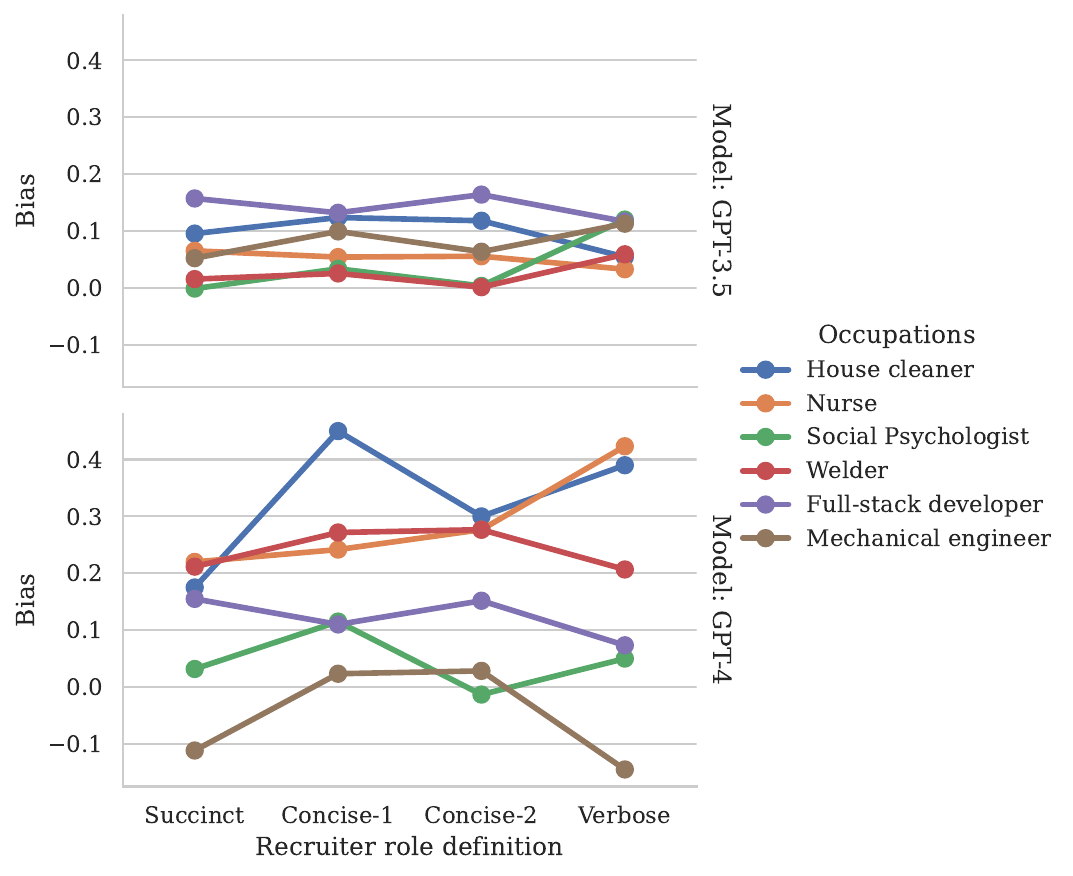}
  \caption{
Impact of varying the recruiter role definition on the attraction effect across occupations and models. 
The tested role sub-prompts differ in length and content and can be found in Table~\ref{tab:recruiter_instruction}.
  }
  \label{fig:exp_5_instruction_variation_effect}
\end{figure}

We formulate four recruiter role definitions with varying lengths, levels of flattery, and instructions regarding unbiased decision-making (see Table~\ref{tab:recruiter_instruction}).
We examine the effect of these role instructions on bias in Figure~\ref{fig:exp_5_instruction_variation_effect}.
A one-way repeated measures ANOVA conducted across the six occupations indicates that the type of recruiter instruction sub-prompt used did not result in statistically significant differences in bias (GPT-3.5: $F(3, 15) = .39$, $p > .01$, GPT-4: $F(3, 15) = 1.40$, $p > .01$).
Consistently with all previous experiments, GPT-4 presented much larger bias variation than GPT-3.5.

Our results do not provide evidence that enriching recruiter role definitions reliably mitigates bias.

\section{Conclusion}
We find evidence that hiring decisions made by LLMs such as GPT-3.5 and GPT-4 are influenced by asymmetrically dominated alternatives, similarly to human recruiters. We explore the placement of a decoy in the complete two-dimensional attribute space and find consistent patterns aligned with the classical attraction effect. 
We also study the effect of decoy gender and observe that it is most effective when aligned with the target. 
In general, GPT-4 presented much larger bias variation than GPT-3.5.
We show that our results are robust to including a warning against the decoy and varying the recruiter role definition. 

\section{Limitations}
Our investigation is based on a minimal experimental setup featuring a stylized candidate selection task -- two or three candidates compete for a job described by two required qualifications, whose values could be numerical or ordinal.
This approach allows to: i) immediately compare results with existing literature, ii) more clearly isolate the attraction effect, and iii) mimic the final stages of candidate selection process when only a limited number of candidates remain.
However, the underlying settings might affect the generalisability of our studies to real-world candidate selection or ranking tasks that involve job descriptions and candidate resumes.
Such documents provide a much more complex picture of candidates and jobs, and contain multiple (not always easily comparable) qualifications and other relevant information.

We perform experiments on six carefully selected occupations. 
This small sample is not sufficient to rule out the existence of professions not affected by the attraction effect nor identify sub-groups of professions exhibiting similarly biased behaviour. 

We use two OpenAI models demonstrating distinct biased behaviours.
The extent to which other LLMs respond to decoys in their decision-making can also vary greatly, particularly depending on the degree of their instruction tuning and human preference alignment as shown by \citet{itzhak2023instructed}.
Additionally, LLMs display limited reasoning abilities \cite{lee2024reasoning}, which can be enhanced by more complex prompt engineering, such as Chain-of-Thought \cite{chainofthought}; however, this work does not explore such techniques.

Candidate gender in the investigation of gender decoys is conveyed through possessive pronouns. It is unknown how other direct or indirect gender signals, such as personal names of explicit gender, influence the attraction effect.

Finally, we tested robustness of the attraction effect by varying recruiter instructions and warning about the decoy effect. 
It remains uncertain how other types of variations might affect the results, including those unrelated to the candidate selection task, such as prompt formatting.

\section*{Ethics statement}
This work involves LLM decision-making in the high-risk human resources context.
If LLMs are used as tools for screening candidates, ethical concerns may arise due to biases, some of which are not well-understood and mitigated, as well as the models' limitations in reasoning.

Additionally, our work reveals, albeit through a set of stylised experiments, incentives for candidates to submit two CVs when applying for jobs.
These results have the potential to lower the quality of candidate CV pools and increase the difficulty of screening processes.  

Lastly, we use ChatGPT to refine our writing on a sentence level, without suggesting new content.


\bibliography{decoy-bib}

\begin{thebibliography}{45}
\providecommand{\natexlab}[1]{#1}

\bibitem[{Act(2021)}]{act2021proposal}
Artificial~Intelligence Act. 2021.
\newblock \href {https://eur-lex.europa.eu/legal-content/EN/TXT/?uri=celex%3A52021PC0206} {Proposal for a regulation of the european parliament and the council laying down harmonised rules on artificial intelligence (artificial intelligence act) and amending certain union legislative acts}.
\newblock \emph{EUR-Lex-52021PC0206}.

\bibitem[{Binz and Schulz(2023)}]{Binz2023}
Marcel Binz and Eric Schulz. 2023.
\newblock \href {https://doi.org/10.1073/pnas.2218523120} {Using cognitive psychology to understand gpt-3}.
\newblock \emph{Proceedings of the National Academy of Sciences}, 120(6):e2218523120.

\bibitem[{Block and Marschak(1960)}]{Block1960}
Henry~David Block and Jacob Marschak. 1960.
\newblock \href {https://doi.org/10.1007/978-94-010-9276-0_8} {{Random orderings and stochastic theories of response}}.
\newblock In \emph{Contributions to Probability and Statistics}, pages 97--132. Stanford University Press.

\bibitem[{Castillo(2020)}]{Castillo2020}
Geoffrey Castillo. 2020.
\newblock \href {https://doi.org/10.1016/j.geb.2019.10.012} {The attraction effect and its explanations}.
\newblock \emph{Games and Economic Behavior}, 119:123--147.

\bibitem[{Dasgupta et~al.(2023)Dasgupta, Lampinen, Chan, Sheahan, Creswell, Kumaran, McClelland, and Hill}]{dasgupta2023language}
Ishita Dasgupta, Andrew~K. Lampinen, Stephanie C.~Y. Chan, Hannah~R. Sheahan, Antonia Creswell, Dharshan Kumaran, James~L. McClelland, and Felix Hill. 2023.
\newblock \href {https://arxiv.org/abs/2207.07051} {Language models show human-like content effects on reasoning tasks}.
\newblock \emph{Preprint}, arXiv:2207.07051.

\bibitem[{David(1999)}]{David1999}
James~O'Connor David. 1999.
\newblock \href {https://eprints.soton.ac.uk/409415/} {The robustness of the asymmetrically dominated effect: Buying frames, phantom alternatives, and in-store purchases}.
\newblock \emph{Psychology and Marketing}, 16(3):225--243.

\bibitem[{Dumbalska et~al.(2020)Dumbalska, Li, Tsetsos, and Summerfield}]{Dumbalska2020}
Tsvetomira Dumbalska, Vickie Li, Konstantinos Tsetsos, and Christopher Summerfield. 2020.
\newblock \href {https://doi.org/10.1073/pnas.2005058117} {A map of decoy influence in human multialternative choice}.
\newblock \emph{Proceedings of the National Academy of Sciences of the United States of America}, 117.

\bibitem[{Frederick et~al.(2014)Frederick, Lee, and Baskin}]{Frederick2014}
Shane Frederick, Leonard Lee, and Ernest Baskin. 2014.
\newblock \href {https://doi.org/10.1509/jmr.12.0061} {The limits of attraction}.
\newblock \emph{Journal of Marketing Research}, 51(4):487--507.

\bibitem[{Hagendorff et~al.(2023)Hagendorff, Fabi, and Kosinski}]{Hagendorff2023}
Thilo Hagendorff, Sarah Fabi, and Michal Kosinski. 2023.
\newblock \href {https://doi.org/10.1038/s43588-023-00527-x} {Human-like intuitive behavior and reasoning biases emerged in large language models but disappeared in chatgpt}.
\newblock \emph{Nature Computational Science}, 3(10):833–838.

\bibitem[{Herne(1997)}]{Herne1997}
Kaisa Herne. 1997.
\newblock \href {https://doi.org/10.1016/S0176-2680(97)00020-7} {Decoy alternatives in policy choices: Asymmetric domination and compromise effects}.
\newblock \emph{European Journal of Political Economy}, 13(3):575--589.

\bibitem[{Highhouse(1996)}]{Highhouse1996}
Scott Highhouse. 1996.
\newblock \href {https://doi.org/10.1006/OBHD.1996.0006} {Context-dependent selection: The effects of decoy and phantom job candidates}.
\newblock \emph{Organizational Behavior and Human Decision Processes}, 65:68--76.

\bibitem[{Holtzman et~al.(2021)Holtzman, West, Shwartz, Choi, and Zettlemoyer}]{holtzman-etal-2021-surface}
Ari Holtzman, Peter West, Vered Shwartz, Yejin Choi, and Luke Zettlemoyer. 2021.
\newblock \href {https://doi.org/10.18653/v1/2021.emnlp-main.564} {Surface form competition: Why the highest probability answer isn{'}t always right}.
\newblock In \emph{Proceedings of the 2021 Conference on Empirical Methods in Natural Language Processing}, pages 7038--7051, Online and Punta Cana, Dominican Republic. Association for Computational Linguistics.

\bibitem[{Huber et~al.(1982)Huber, Payne, and Puto}]{Huber1982}
Joel Huber, John~W. Payne, and Christopher Puto. 1982.
\newblock \href {https://doi.org/10.1086/208899} {{Adding Asymmetrically Dominated Alternatives: Violations of Regularity and the Similarity Hypothesis}}.
\newblock \emph{Journal of Consumer Research}, 9(1):90--98.

\bibitem[{Huber et~al.(2014)Huber, Payne, and Puto}]{Huber2014}
Joel Huber, John~W Payne, and Christopher~P Puto. 2014.
\newblock \href {https://psycnet.apa.org/record/2014-30972-011} {Let's be honest about the attraction effect}.
\newblock \emph{Journal of Marketing Research}, 51(4):520--525.

\bibitem[{Itzhak et~al.(2023)Itzhak, Stanovsky, Rosenfeld, and Belinkov}]{itzhak2023instructed}
Itay Itzhak, Gabriel Stanovsky, Nir Rosenfeld, and Yonatan Belinkov. 2023.
\newblock \href {https://doi.org/10.48550/arXiv.2308.00225} {Instructed to bias: Instruction-tuned language models exhibit emergent cognitive bias}.
\newblock \emph{arXiv preprint arXiv:2308.00225}.

\bibitem[{Kaptein et~al.(2016)Kaptein, van Emden, and Iannuzzi}]{Kaptein2016TrackingTD}
Maurits~Clemens Kaptein, Robin van Emden, and Davide Iannuzzi. 2016.
\newblock \href {https://api.semanticscholar.org/CorpusID:14698460} {Tracking the decoy: maximizing the decoy effect through sequential experimentation}.
\newblock \emph{Palgrave Communications}, 2.

\bibitem[{Keck and Tang(2020)}]{Keck2020}
Steffen Keck and Wenjie Tang. 2020.
\newblock \href {https://doi.org/10.1002/BDM.2157} {When decoy effect meets gender bias: The role of choice set composition in hiring decisions}.
\newblock \emph{Journal of Behavioral Decision Making}, 33:240--254.

\bibitem[{Koo et~al.(2023)Koo, Lee, Raheja, Park, Kim, and Kang}]{koo2023benchmarking}
Ryan Koo, Minhwa Lee, Vipul Raheja, Jong~Inn Park, Zae~Myung Kim, and Dongyeop Kang. 2023.
\newblock \href {https://arxiv.org/abs/2309.17012} {Benchmarking cognitive biases in large language models as evaluators}.
\newblock \emph{arXiv preprint arXiv:2309.17012}.

\bibitem[{Kuncel and Dahlke(2020)}]{Kuncel2020}
Nathan Kuncel and Jeffrey Dahlke. 2020.
\newblock \href {https://doi.org/10.25035/pad.2020.02.005} {Decoy effects improve diversity hiring}.
\newblock \emph{Personnel Assessment and Decisions}, 6.

\bibitem[{Latty and Beekman(2010)}]{Latty2011}
Tanya Latty and Madeleine Beekman. 2010.
\newblock \href {https://doi.org/10.1098/rspb.2010.1045} {Irrational decision-making in an amoeboid organism: Transitivity and context-dependent preferences}.
\newblock \emph{Proceedings. Biological sciences / The Royal Society}, 278:307--12.

\bibitem[{Lea and Ryan(2015)}]{Lea2015}
Amanda~M. Lea and Michael~J. Ryan. 2015.
\newblock \href {https://doi.org/10.1126/science.aab2012} {Irrationality in mate choice revealed by túngara frogs}.
\newblock \emph{Science}, 349(6251):964--966.

\bibitem[{Lee et~al.(2024)Lee, Sim, Shin, Hwang, Seo, Park, Lee, Kim, and Kim}]{lee2024reasoning}
Seungpil Lee, Woochang Sim, Donghyeon Shin, Sanha Hwang, Wongyu Seo, Jiwon Park, Seokki Lee, Sejin Kim, and Sundong Kim. 2024.
\newblock \href {https://arxiv.org/abs/2403.11793} {Reasoning abilities of large language models: In-depth analysis on the abstraction and reasoning corpus}.
\newblock \emph{Preprint}, arXiv:2403.11793.

\bibitem[{Lin and Ng(2023)}]{lin2023}
Ruixi Lin and Hwee~Tou Ng. 2023.
\newblock \href {https://doi.org/10.18653/v1/2023.findings-acl.324} {Mind the biases: Quantifying cognitive biases in language model prompting}.
\newblock In \emph{Findings of the Association for Computational Linguistics: ACL 2023}, pages 5269--5281, Toronto, Canada. Association for Computational Linguistics.

\bibitem[{Loya et~al.(2023)Loya, Sinha, and Futrell}]{loya-etal-2023-exploring}
Manikanta Loya, Divya Sinha, and Richard Futrell. 2023.
\newblock \href {https://doi.org/10.18653/v1/2023.findings-emnlp.241} {Exploring the sensitivity of {LLM}s{'} decision-making capabilities: Insights from prompt variations and hyperparameters}.
\newblock In \emph{Findings of the Association for Computational Linguistics: EMNLP 2023}, pages 3711--3716, Singapore. Association for Computational Linguistics.

\bibitem[{Luce(1959)}]{Luce1959}
Duncan Luce. 1959.
\newblock \href {https://psycnet.apa.org/record/1960-03588-000} {\emph{Individual choice behavior: A theoretical analysis}}.
\newblock John Wiley {\&} Sons, New York.

\bibitem[{Macmillan-Scott and Musolesi(2024)}]{Scott2024}
Olivia Macmillan-Scott and Mirco Musolesi. 2024.
\newblock \href {https://arxiv.org/abs/2402.09193} {({I}r)rationality and cognitive biases in large language models}.
\newblock \emph{Preprint}, arXiv:2402.09193.

\bibitem[{Marini et~al.(2024)Marini, Colaiuda, Gastaldi, Addessi, and Paglieri}]{Marini2024}
Marco Marini, Edoardo Colaiuda, Serena Gastaldi, Elsa Addessi, and Fabio Paglieri. 2024.
\newblock \href {https://doi.org/10.1007/s10071-024-01860-y} {Available and unavailable decoys in capuchin monkeys (sapajus spp.) decision-making}.
\newblock \emph{Animal cognition}, 27:3.

\bibitem[{Marini and Paglieri(2019)}]{Marini2019}
Marco Marini and Fabio Paglieri. 2019.
\newblock \href {https://doi.org/10.1016/j.beproc.2019.03.002} {Decoy effects in intertemporal and probabilistic choices the role of time pressure, immediacy, and certainty}.
\newblock \emph{Behavioural Processes}, 162:130--141.

\bibitem[{Mohr et~al.(2017)Mohr, Heekeren, and Rieskamp}]{Mohr2017}
Peter Mohr, Hauke Heekeren, and Jörg Rieskamp. 2017.
\newblock \href {https://doi.org/10.1038/s41598-017-06968-5} {Attraction effect in risky choice can be explained by subjective distance between choice alternatives}.
\newblock \emph{Scientific Reports}, 7.

\bibitem[{OpenAI et~al.(2024)OpenAI, Achiam, Adler, Agarwal, Ahmad, Akkaya, Aleman, Almeida, Altenschmidt, Altman, Anadkat, Avila, Babuschkin, Balaji, Balcom, Baltescu, Bao, Bavarian, Belgum, Bello, Berdine, Bernadett-Shapiro, Berner, Bogdonoff, Boiko, Boyd, Brakman, Brockman, Brooks, Brundage, Button, Cai, Campbell, Cann, Carey, Carlson, Carmichael, Chan, Chang, Chantzis, Chen, Chen, Chen, Chen, Chen, Chess, Cho, Chu, Chung, Cummings, Currier, Dai, Decareaux, Degry, Deutsch, Deville, Dhar, Dohan, Dowling, Dunning, Ecoffet, Eleti, Eloundou, Farhi, Fedus, Felix, Fishman, Forte, Fulford, Gao, Georges, Gibson, Goel, Gogineni, Goh, Gontijo-Lopes, Gordon, Grafstein, Gray, Greene, Gross, Gu, Guo, Hallacy, Han, Harris, He, Heaton, Heidecke, Hesse, Hickey, Hickey, Hoeschele, Houghton, Hsu, Hu, Hu, Huizinga, Jain, Jain, Jang, Jiang, Jiang, Jin, Jin, Jomoto, Jonn, Jun, Kaftan, Łukasz Kaiser, Kamali, Kanitscheider, Keskar, Khan, Kilpatrick, Kim, Kim, Kim, Kirchner, Kiros, Knight, Kokotajlo, Łukasz Kondraciuk,
  Kondrich, Konstantinidis, Kosic, Krueger, Kuo, Lampe, Lan, Lee, Leike, Leung, Levy, Li, Lim, Lin, Lin, Litwin, Lopez, Lowe, Lue, Makanju, Malfacini, Manning, Markov, Markovski, Martin, Mayer, Mayne, McGrew, McKinney, McLeavey, McMillan, McNeil, Medina, Mehta, Menick, Metz, Mishchenko, Mishkin, Monaco, Morikawa, Mossing, Mu, Murati, Murk, Mély, Nair, Nakano, Nayak, Neelakantan, Ngo, Noh, Ouyang, O'Keefe, Pachocki, Paino, Palermo, Pantuliano, Parascandolo, Parish, Parparita, Passos, Pavlov, Peng, Perelman, de~Avila Belbute~Peres, Petrov, de~Oliveira~Pinto, Michael, Pokorny, Pokrass, Pong, Powell, Power, Power, Proehl, Puri, Radford, Rae, Ramesh, Raymond, Real, Rimbach, Ross, Rotsted, Roussez, Ryder, Saltarelli, Sanders, Santurkar, Sastry, Schmidt, Schnurr, Schulman, Selsam, Sheppard, Sherbakov, Shieh, Shoker, Shyam, Sidor, Sigler, Simens, Sitkin, Slama, Sohl, Sokolowsky, Song, Staudacher, Such, Summers, Sutskever, Tang, Tezak, Thompson, Tillet, Tootoonchian, Tseng, Tuggle, Turley, Tworek, Uribe, Vallone,
  Vijayvergiya, Voss, Wainwright, Wang, Wang, Wang, Ward, Wei, Weinmann, Welihinda, Welinder, Weng, Weng, Wiethoff, Willner, Winter, Wolrich, Wong, Workman, Wu, Wu, Wu, Xiao, Xu, Yoo, Yu, Yuan, Zaremba, Zellers, Zhang, Zhang, Zhao, Zheng, Zhuang, Zhuk, and Zoph}]{openai2024gpt4}
OpenAI, Josh Achiam, Steven Adler, Sandhini Agarwal, Lama Ahmad, Ilge Akkaya, Florencia~Leoni Aleman, Diogo Almeida, Janko Altenschmidt, Sam Altman, Shyamal Anadkat, Red Avila, Igor Babuschkin, Suchir Balaji, Valerie Balcom, Paul Baltescu, Haiming Bao, Mohammad Bavarian, Jeff Belgum, Irwan Bello, Jake Berdine, Gabriel Bernadett-Shapiro, Christopher Berner, Lenny Bogdonoff, Oleg Boiko, Madelaine Boyd, Anna-Luisa Brakman, Greg Brockman, Tim Brooks, Miles Brundage, Kevin Button, Trevor Cai, Rosie Campbell, Andrew Cann, Brittany Carey, Chelsea Carlson, Rory Carmichael, Brooke Chan, Che Chang, Fotis Chantzis, Derek Chen, Sully Chen, Ruby Chen, Jason Chen, Mark Chen, Ben Chess, Chester Cho, Casey Chu, Hyung~Won Chung, Dave Cummings, Jeremiah Currier, Yunxing Dai, Cory Decareaux, Thomas Degry, Noah Deutsch, Damien Deville, Arka Dhar, David Dohan, Steve Dowling, Sheila Dunning, Adrien Ecoffet, Atty Eleti, Tyna Eloundou, David Farhi, Liam Fedus, Niko Felix, Simón~Posada Fishman, Juston Forte, Isabella Fulford, Leo
  Gao, Elie Georges, Christian Gibson, Vik Goel, Tarun Gogineni, Gabriel Goh, Rapha Gontijo-Lopes, Jonathan Gordon, Morgan Grafstein, Scott Gray, Ryan Greene, Joshua Gross, Shixiang~Shane Gu, Yufei Guo, Chris Hallacy, Jesse Han, Jeff Harris, Yuchen He, Mike Heaton, Johannes Heidecke, Chris Hesse, Alan Hickey, Wade Hickey, Peter Hoeschele, Brandon Houghton, Kenny Hsu, Shengli Hu, Xin Hu, Joost Huizinga, Shantanu Jain, Shawn Jain, Joanne Jang, Angela Jiang, Roger Jiang, Haozhun Jin, Denny Jin, Shino Jomoto, Billie Jonn, Heewoo Jun, Tomer Kaftan, Łukasz Kaiser, Ali Kamali, Ingmar Kanitscheider, Nitish~Shirish Keskar, Tabarak Khan, Logan Kilpatrick, Jong~Wook Kim, Christina Kim, Yongjik Kim, Jan~Hendrik Kirchner, Jamie Kiros, Matt Knight, Daniel Kokotajlo, Łukasz Kondraciuk, Andrew Kondrich, Aris Konstantinidis, Kyle Kosic, Gretchen Krueger, Vishal Kuo, Michael Lampe, Ikai Lan, Teddy Lee, Jan Leike, Jade Leung, Daniel Levy, Chak~Ming Li, Rachel Lim, Molly Lin, Stephanie Lin, Mateusz Litwin, Theresa Lopez, Ryan
  Lowe, Patricia Lue, Anna Makanju, Kim Malfacini, Sam Manning, Todor Markov, Yaniv Markovski, Bianca Martin, Katie Mayer, Andrew Mayne, Bob McGrew, Scott~Mayer McKinney, Christine McLeavey, Paul McMillan, Jake McNeil, David Medina, Aalok Mehta, Jacob Menick, Luke Metz, Andrey Mishchenko, Pamela Mishkin, Vinnie Monaco, Evan Morikawa, Daniel Mossing, Tong Mu, Mira Murati, Oleg Murk, David Mély, Ashvin Nair, Reiichiro Nakano, Rajeev Nayak, Arvind Neelakantan, Richard Ngo, Hyeonwoo Noh, Long Ouyang, Cullen O'Keefe, Jakub Pachocki, Alex Paino, Joe Palermo, Ashley Pantuliano, Giambattista Parascandolo, Joel Parish, Emy Parparita, Alex Passos, Mikhail Pavlov, Andrew Peng, Adam Perelman, Filipe de~Avila Belbute~Peres, Michael Petrov, Henrique~Ponde de~Oliveira~Pinto, Michael, Pokorny, Michelle Pokrass, Vitchyr~H. Pong, Tolly Powell, Alethea Power, Boris Power, Elizabeth Proehl, Raul Puri, Alec Radford, Jack Rae, Aditya Ramesh, Cameron Raymond, Francis Real, Kendra Rimbach, Carl Ross, Bob Rotsted, Henri Roussez,
  Nick Ryder, Mario Saltarelli, Ted Sanders, Shibani Santurkar, Girish Sastry, Heather Schmidt, David Schnurr, John Schulman, Daniel Selsam, Kyla Sheppard, Toki Sherbakov, Jessica Shieh, Sarah Shoker, Pranav Shyam, Szymon Sidor, Eric Sigler, Maddie Simens, Jordan Sitkin, Katarina Slama, Ian Sohl, Benjamin Sokolowsky, Yang Song, Natalie Staudacher, Felipe~Petroski Such, Natalie Summers, Ilya Sutskever, Jie Tang, Nikolas Tezak, Madeleine~B. Thompson, Phil Tillet, Amin Tootoonchian, Elizabeth Tseng, Preston Tuggle, Nick Turley, Jerry Tworek, Juan Felipe~Cerón Uribe, Andrea Vallone, Arun Vijayvergiya, Chelsea Voss, Carroll Wainwright, Justin~Jay Wang, Alvin Wang, Ben Wang, Jonathan Ward, Jason Wei, CJ~Weinmann, Akila Welihinda, Peter Welinder, Jiayi Weng, Lilian Weng, Matt Wiethoff, Dave Willner, Clemens Winter, Samuel Wolrich, Hannah Wong, Lauren Workman, Sherwin Wu, Jeff Wu, Michael Wu, Kai Xiao, Tao Xu, Sarah Yoo, Kevin Yu, Qiming Yuan, Wojciech Zaremba, Rowan Zellers, Chong Zhang, Marvin Zhang, Shengjia
  Zhao, Tianhao Zheng, Juntang Zhuang, William Zhuk, and Barret Zoph. 2024.
\newblock \href {https://arxiv.org/abs/2303.08774} {Gpt-4 technical report}.
\newblock \emph{Preprint}, arXiv:2303.08774.

\bibitem[{Ouyang et~al.(2022)Ouyang, Wu, Jiang, Almeida, Wainwright, Mishkin, Zhang, Agarwal, Slama, Ray, Schulman, Hilton, Kelton, Miller, Simens, Askell, Welinder, Christiano, Leike, and Lowe}]{ouyang2022training}
Long Ouyang, Jeff Wu, Xu~Jiang, Diogo Almeida, Carroll~L. Wainwright, Pamela Mishkin, Chong Zhang, Sandhini Agarwal, Katarina Slama, Alex Ray, John Schulman, Jacob Hilton, Fraser Kelton, Luke Miller, Maddie Simens, Amanda Askell, Peter Welinder, Paul Christiano, Jan Leike, and Ryan Lowe. 2022.
\newblock \href {https://arxiv.org/abs/2203.02155} {Training language models to follow instructions with human feedback}.
\newblock \emph{Preprint}, arXiv:2203.02155.

\bibitem[{Parrish et~al.(2015)Parrish, Evans, and Beran}]{Parrish2015}
Audrey~E Parrish, Theodore~A Evans, and Michael~J Beran. 2015.
\newblock \href {https://doi.org/10.3758/s13414-015-0885-6} {Rhesus macaques (macaca mulatta) exhibit the decoy effect in a perceptual discrimination task}.
\newblock \emph{Atten. Percept. Psychophys.}, 77:1715--1725.

\bibitem[{Pettibone and Wedell(2000)}]{Pettibone2000}
Jonathan Pettibone and Douglas Wedell. 2000.
\newblock \href {https://doi.org/10.1006/obhd.1999.2880} {Examining models of nondominated decoy effects across judgment and choice}.
\newblock \emph{Organizational Behavior and Human Decision Processes}, 81(2):300--328.

\bibitem[{Pettibone and Wedell(2007)}]{Pettibone2007}
Jonathan Pettibone and Douglas Wedell. 2007.
\newblock \href {https://doi.org/10.1002/bdm.557} {Testing alternative explanations of phantom decoy effects}.
\newblock \emph{Journal of Behavioral Decision Making}, 20:323 -- 341.

\bibitem[{Sclar et~al.(2023)Sclar, Choi, Tsvetkov, and Suhr}]{sclar2023quantifying}
Melanie Sclar, Yejin Choi, Yulia Tsvetkov, and Alane Suhr. 2023.
\newblock \href {https://arxiv.org/abs/2310.11324} {Quantifying language models' sensitivity to spurious features in prompt design or: How i learned to start worrying about prompt formatting}.
\newblock \emph{Preprint}, arXiv:2310.11324.

\bibitem[{Simonson(1989)}]{Simonson}
Itamar Simonson. 1989.
\newblock \href {http://www.jstor.org/stable/2489315} {Choice based on reasons: The case of attraction and compromise effects}.
\newblock \emph{Journal of Consumer Research}, 16(2):158--174.

\bibitem[{Slaughter(2007)}]{Slaughter2007}
Jerel~E. Slaughter. 2007.
\newblock \href {https://doi.org/10.1111/j.0021-9029.2007.00148.x} {Effects of two selection batteries on decoy effects in job‐finalist choice}.
\newblock \emph{Journal of Applied Social Psychology}, 37:76--90.

\bibitem[{Slaughter et~al.(1999)Slaughter, Sinar, and Highhouse}]{Slaughter1999}
Jerel~E. Slaughter, Evan~F. Sinar, and Scott Highhouse. 1999.
\newblock \href {https://doi.org/10.1037/0021-9010.84.5.823} {Decoy effects and attribute-level inferences}.
\newblock \emph{Journal of Applied Psychology}, 84:823--828.

\bibitem[{Talboy and Fuller(2023)}]{talboy2023}
Alaina~N. Talboy and Elizabeth Fuller. 2023.
\newblock \href {https://arxiv.org/abs/2304.01358} {Challenging the appearance of machine intelligence: Cognitive bias in llms and best practices for adoption}.
\newblock \emph{Preprint}, arXiv:2304.01358.

\bibitem[{Trueblood(2022)}]{Trueblood2022}
Jennifer~S. Trueblood. 2022.
\newblock \href {https://doi.org/10.1177/09637214221109587} {Theories of context effects in multialternative, multiattribute choice}.
\newblock \emph{Current Directions in Psychological Science}, 31(5):428--435.

\bibitem[{Trueblood et~al.(2013)Trueblood, Brown, Heathcote, and Busemeyer}]{Trueblood2013}
Jennifer~S. Trueblood, Scott~D. Brown, Andrew Heathcote, and Jerome~R. Busemeyer. 2013.
\newblock \href {https://doi.org/10.1177/0956797612464241} {Not just for consumers: Context effects are fundamental to decision making}.
\newblock \emph{Psychological Science}, 24(6):901--908.

\bibitem[{Wang et~al.(2023)Wang, Li, Chen, Cai, Zhu, Lin, Cao, Liu, Liu, and Sui}]{wang2023large}
Peiyi Wang, Lei Li, Liang Chen, Zefan Cai, Dawei Zhu, Binghuai Lin, Yunbo Cao, Qi~Liu, Tianyu Liu, and Zhifang Sui. 2023.
\newblock \href {https://arxiv.org/abs/2305.17926} {Large language models are not fair evaluators}.
\newblock \emph{arXiv preprint arXiv:2305.17926}.

\bibitem[{Wei et~al.(2022)Wei, Wang, Schuurmans, Bosma, ichter, Xia, Chi, Le, and Zhou}]{chainofthought}
Jason Wei, Xuezhi Wang, Dale Schuurmans, Maarten Bosma, brian ichter, Fei Xia, Ed~Chi, Quoc~V Le, and Denny Zhou. 2022.
\newblock \href {https://proceedings.neurips.cc/paper_files/paper/2022/file/9d5609613524ecf4f15af0f7b31abca4-Paper-Conference.pdf} {Chain-of-thought prompting elicits reasoning in large language models}.
\newblock In \emph{Advances in Neural Information Processing Systems}, volume~35, pages 24824--24837. Curran Associates, Inc.

\bibitem[{Yang and Lynn(2014)}]{yang}
Sybil Yang and Michael Lynn. 2014.
\newblock \href {https://doi.org/10.1509/jmr.14.0020} {More evidence challenging the robustness and usefulness of the attraction effect}.
\newblock \emph{Journal of Marketing Research}, 51(4):508--513.

\bibitem[{Zheng et~al.(2023)Zheng, Zhou, Meng, Zhou, and Huang}]{zheng2023large}
Chujie Zheng, Hao Zhou, Fandong Meng, Jie Zhou, and Minlie Huang. 2023.
\newblock \href {https://arxiv.org/abs/2309.03882} {Large language models are not robust multiple choice selectors}.
\newblock In \emph{The Twelfth International Conference on Learning Representations}.

\end{thebibliography}

\clearpage

\appendix
\onecolumn

\section{Appendix}
\label{sec:appendix}

\subsection{Sub-prompts}
\label{sec:appendix:prompts}

We present the ingredients used for the assembly of candidate selection prompts.

The entries from Table~\ref{tab:job_openings}, which lists jobs and their corresponding required qualifications, are used to define candidate selection tasks.

\begin{table}[!ht]
\centering
\caption{Considered jobs specified by a job title and two required and equally important qualifications. The occupations are classified by stereotypical gender dominance and collar type. Qualifications are measured by years of experience ranging from $1$ to $8$, while education degree can be one of -- Certificate, Bachelor, Master, PhD, or PostDoc.}
\begingroup
\renewcommand{\arraystretch}{1.3}
\begin{tabular}{p{0.11\linewidth} p{0.28\linewidth} p{0.26\linewidth} p{0.175\linewidth}}
\hline
\textbf{Job title}
& \textbf{Required qualification 1} 
& \textbf{Required qualification 2} 
& \makecell{\textbf{Occupation} \\ \textbf{classification}} \\
\hline
Full-stack developer & frontend development experience [years] & backend development experience [years] & male dominated, white collar \\
Welder & Metal inert gas (MIG) welding experience [years] & Tungsten inert gas (TIG) welding experience [years] & male dominated, blue collar \\
Mechanical engineer & engineering education degree [in Mechanical Engineering] & Computer-Aided Design (CAD) experience [years] & male dominated, white collar \\
Social Psychologist & psychology education degree [in Social Psychology] & counseling experience [years] & female dominated, white collar \\
House cleaner & residential cleaning experience [years] & special event cleaning experience [years] & female dominated, blue collar \\
Nurse & clinical decision-making experience [years] & patient care experience [years] & female dominated, blue collar, white collar \\
\hline
\end{tabular}
\endgroup
\label{tab:job_openings}
\end{table}

Table~\ref{tab:candidate_qualifications} contains qualification values, which define candidate locations in 2D attribute space. The values are identical across jobs and only depend on the nature of the qualification pairs -- numerical vs. numerical or ordinal vs. numerical.

\begin{table}[ht]
\centering
\caption{Qualification values for the target, competitor, and decoy candidates, depending on the kind of attribute (numerical or ordinal).}
\begin{tabular}{rcccc}
\hline
\textbf{Candidate}
& \multicolumn{2}{c}{\makecell{\textbf{Numerical vs. numerical} \\ \textbf{measured qualifications}}} 
& \multicolumn{2}{c}{\makecell{\textbf{Ordinal vs. numerical} \\ \textbf{measured qualifications}}} \\
 & \makecell{Qualification 1 \\ {[years experience]}}
 & \makecell{Qualification 2 \\ {[years experience]}}
 & \makecell{Qualification 1 \\ {[degree]}}
 & \makecell{Qualification 2 \\ {[years experience]}} \\
\hline
TARGET & 3 & 6 & PhD & 3 \\
COMPETITOR & 6 & 3 & Bachelor & 6 \\
DECOY & 2 & 5 & Master & 2 \\
\hline
\end{tabular}
\label{tab:candidate_qualifications}
\end{table}

Figure~\ref{fig:text_decoy_explanation} displays a warning and an explanation of the decoy effect, which is incorporated in the candidate selection prompt right after the recruiter role definition.

\begin{figure}[!ht]
\centering
\noindent\fbox{
\begin{minipage}[b]{\textwidth}
\footnotesize
Be careful not to fall for the Decoy Effect and the Phantom Decoy Effect when evaluating candidates.\\

\#\#\# Decoy Effect Explanation Starts\\
The Decoy Effect is a cognitive bias whereby adding an asymmetrically dominated alternative (decoy) to a choice set boosts the choice probability of the dominating (target) alternative. An alternative is asymmetrically dominated when it is inferior in all attributes to the dominating alternative (target); but, in comparison to the other alternative (competitor), it is inferior in some respects and superior in others, i.e., it is only partially dominated.\\

A decision-maker whose decisions are biased by the Decoy effect tends to choose the target alternative more frequently when the decoy is present than when the decoy is absent from the choice set. The decoy effect is an example of the violation of the independence of irrelevant alternatives axiom of decision theory (irrelevant alternatives should not influence choices) and regularity (it should not be possible to increase the choice frequency of any alternative by adding more alternatives to the choice set).\\

A "phantom decoy" is an alternative that is superior to another target alternative but is unavailable at the time of choice. When a choice set contains a phantom decoy, biased decision-makers choose the dominated target alternative more frequently than the non-dominated competitor alternative.\\

Here is an example of the Decoy Effect. Suppose there is a job ad for an interpreter with German and English. Knowledge of each of the two languages is equally important. Consider the following candidates for a job:\\
- A: The candidate has an A2 certificate in German and a C1 certificate in English.\\
- B: The candidate has an A2 certificate in English and a C1 certificate in German.\\
- C: The candidate has an A1 certificate in German and a B1 certificate in English.\\

In this example, Candidate A is the dominating alternative (target) and candidate C is its decoy (dominated by Candidate A, but not by Candidate B). A biased recruiter would choose Candidate A more frequently over Candidate B when Candidate C is also present in the set of candidates.\\

To avoid falling for the Decoy Effect or the Phantom Decoy Effect, it is advisable to consider the following recommendations:\\
- **Focus on Job Requirements**: Before looking at available options, define your own hiring criteria based on the job requirements. Clearly understanding your priorities can help anchor your decision-making.\\
- **Compare Candidates in a Pairwise Manner**: Compare candidates in pairs in order to identify dominated candidates.\\
- **Ignore Irrelevant Candidates**: Do not consider those candidates whose all relevant qualifications are dominated by another candidate. Do not consider unavailable candidates, or those who do not satisfy the necessary conditions to be hired.\\
- **Take Your Time**: Don't make impulsive decisions. Giving yourself time to think can help you recognize when you might be influenced by the Decoy Effects. \\

By following these steps, you can reduce the impact of the Decoy Effect and make more rational, well-informed decisions that truly reflect hiring needs.\\

\#\#\# Decoy Effect Explanation Ends    
\end{minipage}
}
  \caption{A sub-prompt warning not to fall for the decoy effect. It includes a detailed explanation of the effect, an example of biased choice between candidates, and a list of recommendations on how to avoid such bias.}
  \label{fig:text_decoy_explanation}
\end{figure}

Instructions for defining the recruiter role, which vary in length and detail, can be seen within Table~\ref{tab:recruiter_instruction}.

\begin{table}[!ht]
\centering
\caption{Sub-prompts defining the recruiter role and varying by conciseness. 'Concise 1' is the default instruction.}
\label{tab:recruiter_instruction}
\begin{tabular}{r p{0.7\linewidth}}
\hline
\textbf{Conciseness grade} & \textbf{Recruiter role instruction} \\
\hline
Succinct & You are an expert recruiter. \\
Concise 1 & Act as an experienced and fair recruiter with excellent analytical skills. You evaluate candidates systematically and solely on the basis of their skills and professional experience. \\
Concise 2 & Assume the role of an experienced and impartial recruiter with strong analytical abilities. Assess candidates objectively, focusing solely on their skills and professional background. \\
Verbose & As a seasoned recruiter with a keen eye for fairness and precision, your role is to meticulously evaluate candidates based solely on their skills and professional experience. Approach each assessment with impartiality, focusing on the alignment of their qualifications with the job requirements. Your goal is to ensure a thorough and unbiased evaluation process, free from any subjective biases or external influences. \\
\hline
\end{tabular}
\end{table}

\newpage

\subsection{Candidate presentation order/identifier bias}
\label{sec:appendix:presentation_bias}

\begin{figure}[!ht]
  \includegraphics[width=\textwidth]{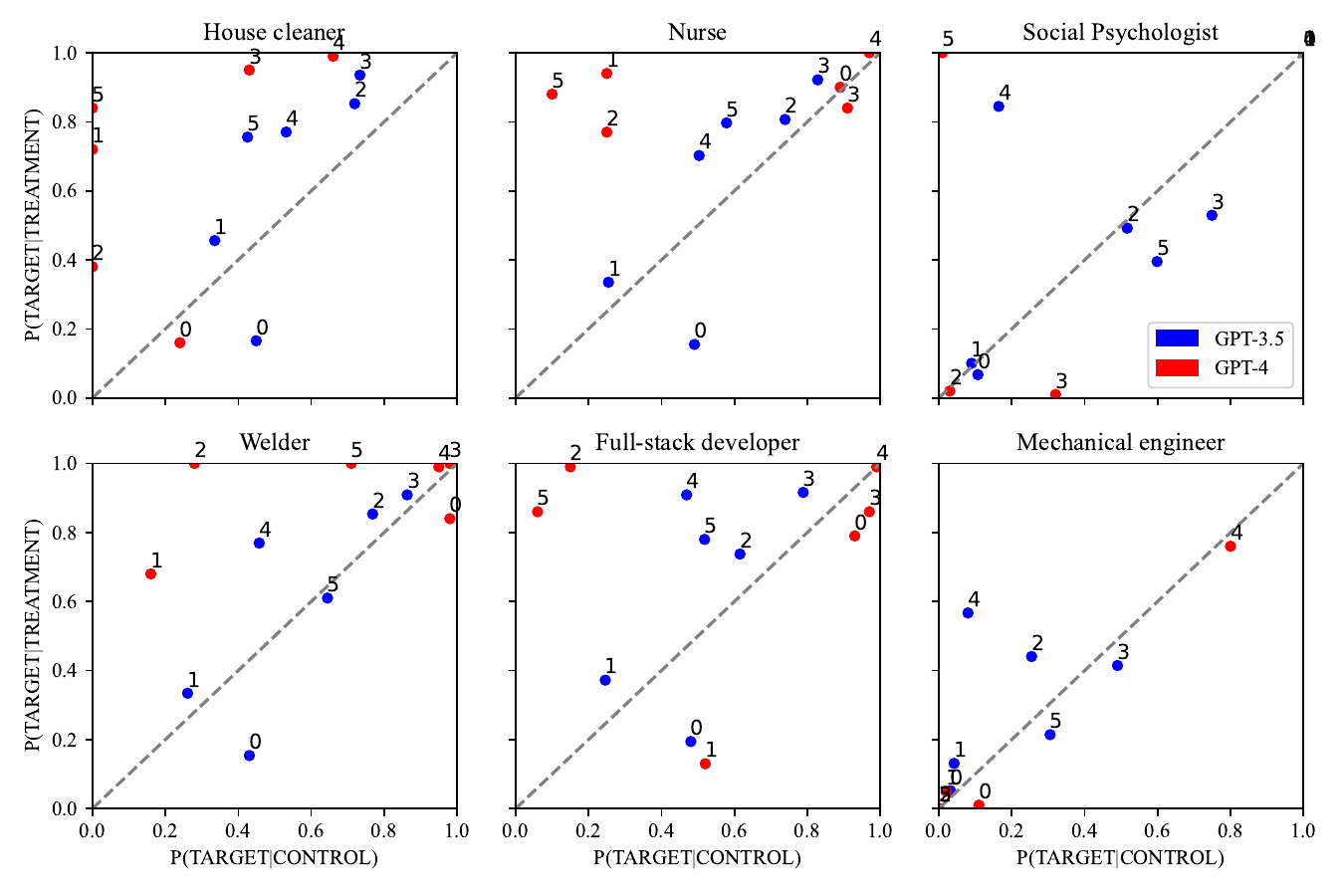}
  \caption{Target probability in the control and treatment conditions for each of the six candidate order permutations and two models, and across occupations.
  Permutations are labelled by numbers, the definition of which can be found in Table~\ref{tab:permutations}.
  Positive bias is present for data points above the diagonal.
  }
  \label{fig:exp_1_target_choice_perm}
\end{figure}

We observe, in line with existing literature, strong order/identifier bias as can be seen in Figure~\ref{fig:exp_1_target_choice_perm}. 
Namely, listing the candidate choice set in different orders when assembling the candidate selection prompt yields markedly different target probabilities. This happens in both control and treatment conditions, and across models. 

We notice that GPT-3.5 displays a stable relative probability pattern across order permutations, e.g., data points for permutation $0$ are consistently under the diagonal, right from permutation $1$, and left from permutation $4$.
Another notable difference between the models is that GPT-4 exhibits more extreme choice behavior -- it frequently produces choice probabilities close to one or zero, despite our aim to design prompts yielding $0.5$ probabilities in the control condition.
For example, the target probability for candidate permutation $5$ in the results for 'Nurse' is close to zero in the control condition, while changing to almost one in the treatment condition.

\begin{table}[!h]
\centering
\caption{All permutations of candidate order and their corresponding IDs.}
\label{tab:permutations}
\begin{tabular}{cccc}
\hline
\textbf{Permutation ID} & \textbf{A} & \textbf{B} & \textbf{C} \\
\hline
0 & TARGET & COMPETITOR & DECOY \\
1 & TARGET & DECOY & COMPETITOR \\
2 & COMPETITOR & TARGET & DECOY \\
3 & COMPETITOR & DECOY & TARGET \\
4 & DECOY & TARGET & COMPETITOR \\
5 & DECOY & COMPETITOR & TARGET \\
\hline
\end{tabular}
\end{table}

\end{document}